\documentclass{aastex62}

\usepackage{amssymb}
\usepackage{latexsym}
\usepackage{amsmath}
\usepackage{color}
\usepackage{graphicx,dblfloatfix}
\usepackage{subfigure}

\newcommand{\bfm}[1]{\mbox{\boldmath{$#1$}}}

\graphicspath{{./}{figures/}}

\submitjournal{ApJ Letters}

\shorttitle{Rotationally induced failure of Ryugu}
\shortauthors{Hirabayashi et al.}

\begin{document}

\title{The western bulge of 162173 Ryugu formed as a result of a rotationally driven deformation process}

\correspondingauthor{Masatoshi Hirabayashi}
\email{thirabayashi@auburn.edu}

\author{Masatoshi Hirabayashi}
\affiliation{Auburn University, 211 Davis Hall, Auburn, Alabama, 36849, United States}

\author{Eri Tatsumi}
\affiliation{University of Tokyo, 7-3-1 Hongo, Bunkyo-ku, Tokyo 113-0033, Japan}

\author{Hideaki Miyamoto}
\affiliation{University of Tokyo, 7-3-1 Hongo, Bunkyo-ku, Tokyo 113-8656, Japan}

\author{Goro Komatsu}
\affiliation{Universit\`{a} degli Studi ``G. d'Annunzio", Viale Pindaro 42, 65127 Pescara, Italy}

\author{Seiji Sugita}
\affiliation{University of Tokyo, 7-3-1 Hongo, Bunkyo-ku, Tokyo 113-0033, Japan}

\author{Sei-ichiro Watanabe}
\affiliation{Nagoya University, Furo-cho, Chikusa-ku, Nagoya, Aichi 464-8601, Japan}

\author{Daniel J. Scheeres}
\affiliation{University of Colorado Boulder, 429 UCB, Boulder, Colorado 80309 United States}

\author{Olivier S. Barnouin}
\affiliation{Applied Physics Laboratory/Johns Hopkins University, 11100 Johns Hopkins Road, Laurel, Maryland 20723, United States}

\author{Patrick Michel}
\affiliation{Universit{\'e} C{\^o}te d'Azur, Observatoire de la C{\^o}te d'Azur, CNRS, Laboratoire Lagrange, CS 34229, 06304 Nice Cedex 4, France}

\author{Chikatoshi Honda}
\affiliation{University of Aizu, Ikki-machi, Aizuwakamatsu, Fukushima 965-8580, Japan}

\author{Tatsuhiro Michikami}
\affiliation{Kindai University, 1 Takaya Umenobe, Higashi-Hiroshima, Hiroshima 739-2116}

\author{Yuichiro Cho}
\affiliation{University of Tokyo, 7-3-1 Hongo, Bunkyo-ku, Tokyo 113-0033, Japan}

\author{Tomokatsu Morota}
\affiliation{Nagoya University, Furo-cho, Chikusa-ku, Nagoya, Aichi 464-8601, Japan}

\author{Naru Hirata}
\affiliation{University of Aizu, Ikki-machi, Aizuwakamatsu, Fukushima 965-8580, Japan}

\author{Naoyuki Hirata}
\affiliation{Kobe University, 1-1 Rokkodaicho, Nada, Kobe, Hyogo 657-8501, Japan}

\author{Naoya Sakatani}
\affiliation{Institute of Space and Astronautical Science, Japan Aerospace Exploration Agency, 3-1-1 Yoshinodai, Chuo-ku, Sagamihara, Kanagawa 252-5210, Japan}

\author{Stephen R. Schwartz}
\affiliation{University of Arizona, 1629 E University Blvd, Tucson, Arizona 85721, United States}

\author{Rie Honda}
\affiliation{Kochi University, Akebono-Cho, Kochi, Kochi 780-8520, Japan}

\author{Yasuhiro Yokota}
\affiliation{Kochi University, Akebono-Cho, Kochi, Kochi 780-8520, Japan}
\affiliation{Institute of Space and Astronautical Science, Japan Aerospace Exploration Agency, 3-1-1 Yoshinodai, Chuo-ku, Sagamihara, Kanagawa 252-5210, Japan}

\author{Shingo Kameda}
\affiliation{Rikkyo University, 3-34-1 Nishi-Ikebukuro,Toshima-ku, Tokyo 171-8501, Japan}

\author{Hidehiko Suzuki}
\affiliation{Meiji University, 1-1-1 Higashi-Mita, Tama-ku, Kawasaki, Kanagawa 214-8571, Japan} 

\author{Toru Kouyama}
\affiliation{National Institute of Advanced Industrial Science and Technology, 1-1-1 Umezono, Tsukuba, Ibaraki 305-8560 Japan}

\author{Masahiko Hayakawa}
\affiliation{Institute of Space and Astronautical Science, Japan Aerospace Exploration Agency, 3-1-1 Yoshinodai, Chuo-ku, Sagamihara, Kanagawa 252-5210, Japan}

\author{Moe Matsuoka}
\affiliation{Institute of Space and Astronautical Science, Japan Aerospace Exploration Agency, 3-1-1 Yoshinodai, Chuo-ku, Sagamihara, Kanagawa 252-5210, Japan}

\author{Kazuo Yoshioka}
\affiliation{University of Tokyo, 7-3-1 Hongo, Bunkyo-ku, Tokyo 113-0033, Japan}

\author{Kazunori Ogawa}
\affiliation{Kobe University, 1-1 Rokkodaicho, Nada, Kobe, Hyogo 657-8501, Japan}

\author{Hirotaka Sawada}
\affiliation{Institute of Space and Astronautical Science, Japan Aerospace Exploration Agency, 3-1-1 Yoshinodai, Chuo-ku, Sagamihara, Kanagawa 252-5210, Japan}

\author{Makoto Yoshikawa}
\affiliation{Institute of Space and Astronautical Science, Japan Aerospace Exploration Agency, 3-1-1 Yoshinodai, Chuo-ku, Sagamihara, Kanagawa 252-5210, Japan}

\author{Yuichi Tsuda}
\affiliation{Institute of Space and Astronautical Science, Japan Aerospace Exploration Agency, 3-1-1 Yoshinodai, Chuo-ku, Sagamihara, Kanagawa 252-5210, Japan}



\begin{abstract}
162173 Ryugu, the target of Hayabusa2, has a round shape with an equatorial ridge, which is known as a spinning top-shape. A strong centrifugal force is a likely contributor to Ryugu's top-shaped features. Observations by Optical Navigation Camera onboard Hayabusa2 show a unique longitudinal variation in geomorphology; the western side of this asteroid, later called the western bulge, has a smooth surface and a sharp equatorial ridge, compared to the other side. Here, we propose a structural deformation process that generated the western bulge. Applying the mission-derived shape model, we employ a finite element model technique to analyze the locations that experience structural failure within the present shape. Assuming that materials are uniformly distributed, our model shows the longitudinal variation in structurally failed regions when the spin period is shorter than $\sim3.75$ h. Ryugu is structurally intact in the subsurface region of the western bulge while other regions are sensitive to structural failure. We infer that this variation is indicative of the deformation process that occurred in the past, and the western bulge is more relaxed structurally than the other region. Our analysis also shows that this deformation process might occur at a spin period between $\sim3.5$ h and $\sim3.0$ h, providing the cohesive strength ranging between $\sim4$ Pa and $\sim 10$ Pa.  
\end{abstract}



\section{Introduction} \label{sec:intro}
When an asteroid has a round shape with equatorial ridges, this asteroid is called a spinning top-shaped asteroid. Ground radar observations have shown that spinning top-shapes are common in the solar system \citep{Benner2015}. 162173 Ryugu \citep{Watanabe2019} and 101955 Bennu \citep{Scheeres2016Bennu, Scheeres2019Bennu} are examples of spinning top-shapes. Such asteroids may have natural satellites \citep[e.g.][]{Ostro2006, Michel2016, Naidu2016, Brozovic2011, Fang2011, Becker2015}. Importantly, a spinning top-shape is found to be independent of material composition \citep{Benner2015}.

The formation of a spinning top-shaped asteroid may result from a reaccumulation process after catastrophic disruption \citep{Michel2018, Michel2019} and/or quasi-static spin-up of a spheroidal body driven by micrometeorite impacts or a solar radiation pressure-driven torque, which is the so-called YORP effect \citep{Rubincam2000}. Regarding the quasi-static spin-up process, there are two spinning top-shape formation scenarios depending on the internal strength distribution: surface mass movement if there is a strong interior covered by a weak layer \citep{Walsh2008, Walsh2012} and internal deformation if there is uniform structure \citep{Hirabayashi2014DA, Hirabayashi2015internal, Zhang2017, Hirabayashi2018}. Note that even if the rotational period is relatively long, the uniform structure may also cause surface failure \citep{Hirabayashi2015Sphere}. All these processes make the shape settle into the minimum potential configuration \citep{Scheeres2015Land, Walsh2018}. 

\begin{figure}[ht]
  \centering
  \includegraphics{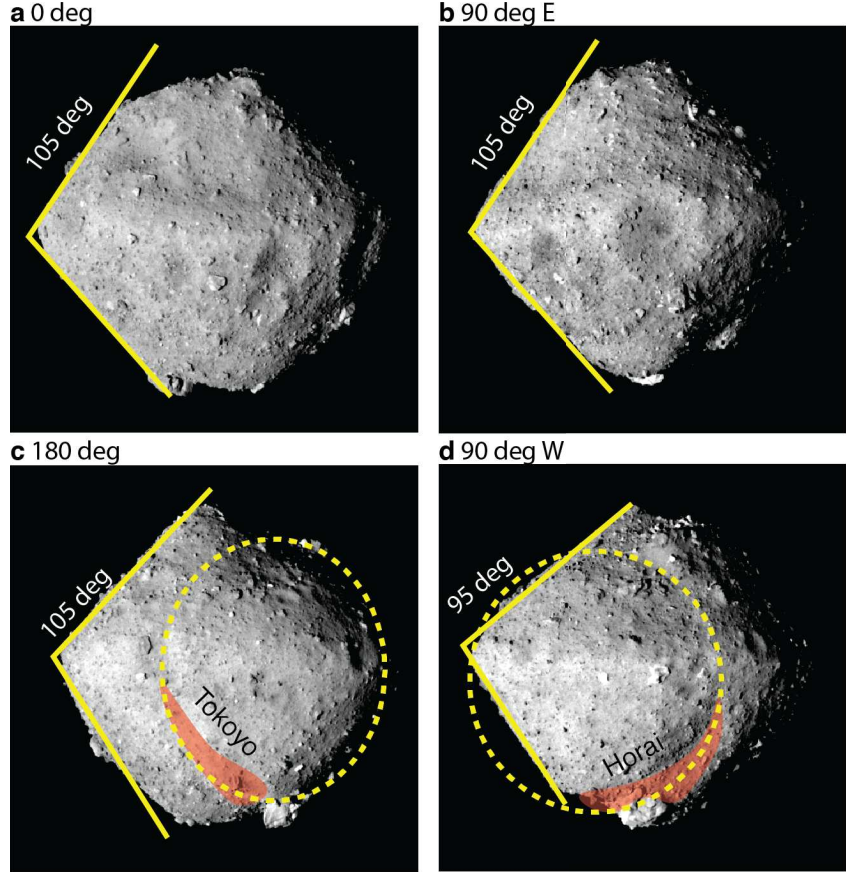}
  \caption{Ryugu seen from the Hayabusa2 spacecraft at different views. a. 0 deg (image: hyb2\_onc\_20180630\_074839\_tvf\_l2a). b. 90 deg East (image: hyb2\_onc\_20180630\_140543\_tvf\_l2a). c. 180 deg (image: hyb2\_onc\_20180630\_120959\_tvf\_l2a). d. 90 deg West (image: hyb2\_onc\_20180630\_101759\_tvf\_l2a). The yellow circles indicate the western bulge. The red regions are Tokoyo and Horai Fossae. The ridge angles are determined by only considering the sunlit sides and by averaging the slopes at the middle latitudes. Image credit: JAXA/UTokyo/Kochi U./Rikkyo U./Nagoya U./Chiba Ins. Tech/Meiji U./U. Aizu/AIST.} 
  \label{Fig:shape}  
\end{figure}

Ryugu is the target of the Hayabusa2 mission led by JAXA \citep{Watanabe2019}. Figure \ref{Fig:shape} describes images taken by the telescope camera, which is part of the Optical Navigation Camera (ONC) system, onboard the spacecraft \citep{Kameda2017, Suzuki2018, Tatsumi2018}. This asteroid has the common shape features of a spinning top-shaped asteroid, and its spin axis is perpendicular to the equatorial ridge \citep{Watanabe2019}. The currently reported spin period is 7.63262 h \citep{Watanabe2019}, which is much longer than the $\sim2.3$-h spin barrier, the typical spin limit of small rubble pile asteroids \citep{Pravec2008}. The bulk density was estimated as $\sim$1.19 g cm$^{-3}$ from the orbital motion of the Hayabusa2 spacecraft \citep{Watanabe2019}. 

The surface condition of Ryugu was observed to be longitudinally divided into the eastern and western regions \citep{Sugita2019}. These areas are distinguished by Tokoyo and Horai Fossae, a trough system that is widely placed in the southern hemisphere and possibly spreads towards the northern hemisphere (Figure \ref{Fig:shape}). The western region (160 deg E - 70 deg W), later known as the western bulge, is apparently less affected by crater bombardment and thus is smoother than the other side \citep{Sugita2019}. Furthermore, while Ryugu's spinning top-shape is axisymmetric in general; however, it is not strictly so (Figure \ref{Fig:shape}). The ridge angle, which defines an angle between the surfaces in the northern and southern hemispheres at the equatorial ridge, or Ryujin Dorsum, varies when seen from different views. The ridge angle of the western bulge is 95 deg (Panel d in Figure \ref{Fig:shape}) and is sharper than that of other sides, which is 105 deg (Panels a through c in Figure \ref{Fig:shape}). This feature was also discussed in \cite{Watanabe2019} in detail. 

This study tests a hypothesis that this longitudinal dichotomy results from shape deformation when Ryugu was rotating with a short rotational period in the past. Following is the outline of this paper. First, we employ the finite element model (FEM) analysis \citep{Hirabayashi2018} by assuming that the material distribution is uniform, which is compatible with observations \citep{Sugita2019, Watanabe2019}. Second, we use a dynamical analysis model to determine at what spin period Ryugu would have experienced the shape deformation process that caused the western bulge. Our work extends the structural analysis by \cite{Watanabe2019}, who only focused on the general internal deformation process at spin periods of 3.75 hr and 3.5 hr, to analyze the deformation process that implies the formation of the western bulge at fast rotation. 

\section{How did Ryugu deform at a short spin period?}
We employ a FEM analysis \citep{Hirabayashi2016Nature, Hirabayashi2018} to investigate the structural failure mode of Ryugu at a given spin period. This technique is the same as that done by \cite{Watanabe2019}. In this work, we use the shape model developed from the stereophotoclinometry (SPC) technique \citep{Watanabe2019}. The version used is SHAPE\_SPC\_3M\_v20180731.obj in which there are 1,579,015 vertices and 3,145,728 facets \citep{Watanabe2019}. To avoid computational intensity, we reduce the size of the shape model by using the Quadric Edge Collapse Decimation technique available in Meshlab (http://www.meshlab.net/). The newly derived shape model consists of 3,072 facets and 1,538 vertices.

We use a FEM solver available in ANSYS Mechanical APDL (18.1) under the license owned by Auburn University's Samuel Ginn College of Engineering. We develop a three-dimensional, 10-node FEM mesh for the current shape of Ryugu by using TetGen, an open tetrahedron-mesh generator \citep{Si2015}, and a mesh-refining tool in ANSYS \citep{Hirabayashi2018}. The generated FEM mesh consists of 8,308 elements and 15,038 nodes. This model assumes that spin-up acceleration is so small that dynamic change in stress is negligible. Thus, we solve the equilibrium equation of structure: 
\begin{eqnarray}
\frac{\partial \sigma_{ij}}{\partial x_j} + \rho b_i = 0. \label{Eq:equilibrium}
\end{eqnarray}
where $\rho$ is the bulk density, $\sigma_{ij}$ is the stress component ($i,j = 1,2,3$), and $b_j$ is the body force. Also, we use the Einstein notation in this equation. For elastic deformation, we use a linear-elastic rule, i.e., Hookean deformation. To describe inelastic deformation, we assume a plastic flow to be associated with a yield criterion where the behavior of materials becomes inelastic (the associate flow rule) \citep{Hirabayashi2016Nature}. A plastic flow is given as a function of a partial derivative of a yield criterion with respect to stress \citep{Chen1988}. To model the yield condition of regolith in Ryugu, we use the Drucker-Prager yield criterion, which depends on the cohesive strength and the angle of friction. In this work, the angle of friction is fixed at 35$^\circ$, which is a typical value for regolith and soil materials \citep{Lambe1969}. On the other hand, we consider the cohesive strength to be a free parameter to investigate the failure mode at a given spin period \citep{Hirabayashi2018}. The bulk density is fixed at 1.2 g cm$^{-3}$. 

We introduce two critical parameters in our analysis. The first parameter is the smallest cohesive strength that can hold the present shape, which is later known as the minimum cohesive strength \citep{Hirabayashi2018}. This parameter is an ideal value that could induce structural failure somewhere in Ryugu's body. Thus, if the actual cohesive strength is lower than the minimum cohesive strength, structural failure should happen, causing deformation. We describe the minimum cohesive strength as a function of the spin period, which is called a failure mode diagram (FMD) \citep{Hirabayashi2018}. The second parameter is the stress ratio, a ratio of the stress state to the yield stress. If an element has a stress ratio reaching unity, it fails structurally.

\begin{figure}[ht]
  \centering
  \includegraphics{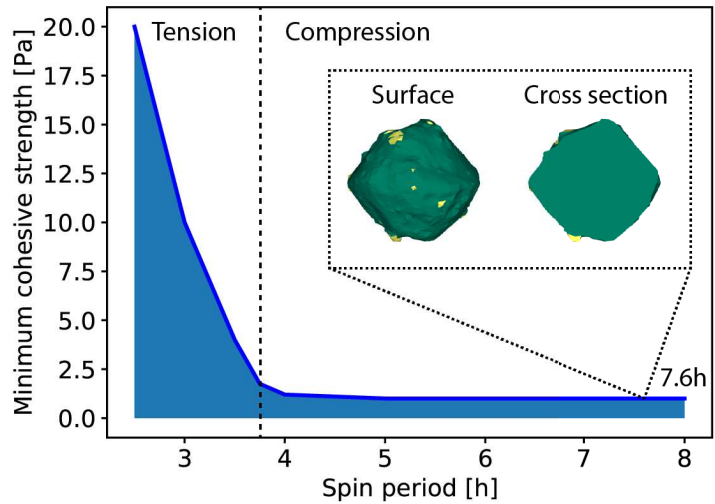}
  \caption{Failure mode diagram of Ryugu. The $x$ axis is the spin period, while the $y$ axis is the minimum cohesive strength. The dashed line indicates the boundary between compression-dominant deformation and tension-dominant deformation at a spin period of 3.75 h. The shaded region indicates that Ryugu cannot structurally exist because the cohesive strength is below the minimum cohesive strength. The top right panel describes the failure mode of Ryugu at a spin period of 7.6 hr. The green and yellow regions are structurally intact and failed areas, respectively.} 
  \label{Fig:failedHour}  
\end{figure}

We compute the failure mode of the present shape at nine spin periods: 2.5 h, 3.0 h, 3.5 h, 3.75 h, 4.0 h, 5.0 h, 6.0 h, 7.6 h, and 8.0 h. Note that \cite{Watanabe2019} only introduced the cases at spin periods of 3.5 h and 3.75 h. Figure \ref{Fig:failedHour} shows the FMD of Ryugu. At a spin period longer than 3.75 h, inelastic deformation may appear due to shear stress under compression if materials are nearly cohesionless. Such sensitive regions depend on local topography and are located in small regions on the surface (Figure \ref{Fig:failedHour}). Thus, we rule out the scenario that inelastic deformation at slow rotation produced the western bulge. At fast rotation, on the other hand, because the centrifugal force plays a dominant role in the stress distribution, Ryugu needs high cohesive strength to hold the present shape. Failed regions start to appear in the interior of the body when the spin period is shorter than 3.75 hr \citep{Watanabe2019}. Also, the minimum cohesive strength monotonically increases as the spin period becomes short (Figure \ref{Fig:failedHour}). 

In this paper, to discuss the deformation process that made the western bulge, we employ the following approach. Our technique only accounts for the present shape and provides the failure mode that this shape experiences, meaning that we do not directly show the failure mode that the precursor shape had. To infer the deformation process in the past, we search for structurally intact points in the present shape. When a deformation process occurs at a given spin period, the deformed area may settle into a configuration that the potential of the shape becomes minimum \citep{Scheeres2015Land, Walsh2018}. Therefore, when Ryugu encounters a fast-rotation condition once again, only the elements that did not experience structural failure previously become sensitive to failure. The originally failed regions, on the other hand, may not experience structural failure because they are structurally relaxed.  

\begin{figure}[ht]
  \centering
  \includegraphics{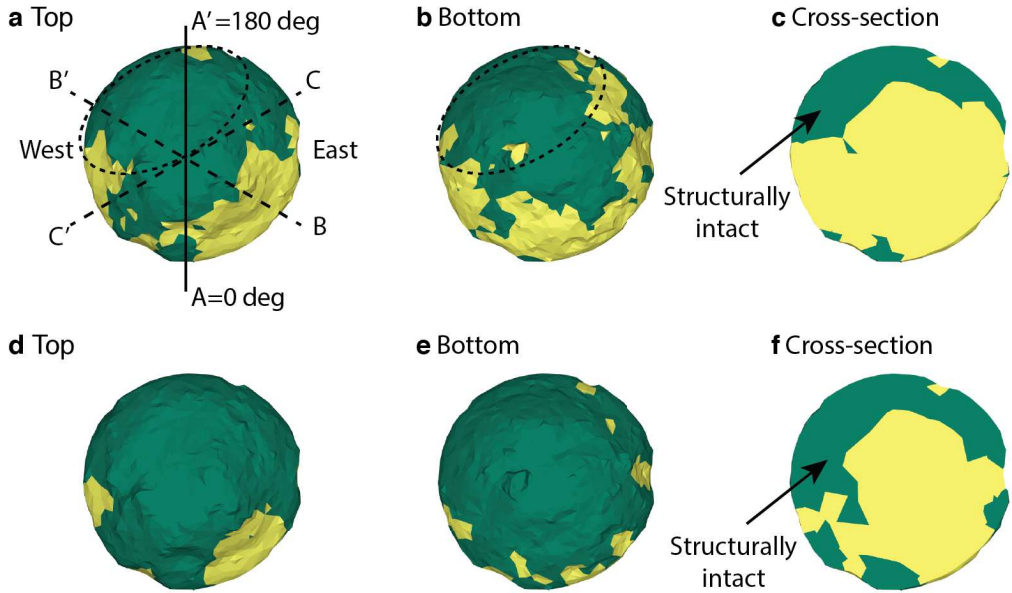}
  \caption{Distributions of the failed regions (yellow) at spin periods of 3.75 h (top) and 3.5 h (bottom). The left, middle, and right columns show the top surface, the bottom surface, and the cross-section, respectively. The circle with a dashed line corresponds to the western bulge \citep{Sugita2019}. Lines $AA^\prime$, $BB^\prime$, and $CC^\prime$ define the cross-sections, which are cut through longitudes of 0$^\circ$ and 180$^\circ$, 60$^\circ$ E and 120$^\circ$ W, and 120$^\circ$ E and 60$^\circ$ W.} 
  \label{Fig:top_bottom}  
\end{figure}

If the body is axisymmetric, the deformation process should be symmetric because the centrifugal and gravity forces are the main loadings. If an asteroid is usually irregular in shape, the failure mode becomes complex and asymmetric \citep{Hirabayashi2018}. Our FEM results show that the structurally intact region corresponds to the western bulge. Figure \ref{Fig:top_bottom} gives the failed regions viewed from the spin axis direction. Lines $AA^\prime$, $BB^\prime$, and $CC^\prime$ are the cross sections at three different longitudes. Lines $AA^\prime$ and $BB^\prime$ cross the western bulge while Line $CC^\prime$ does not. While the equatorial plane and surface widely experience structural failure at spin periods of both 3.75 h and 3.5 h, the structurally intact regions appear at longitudes between 120$^\circ$ E and 90$^\circ$ W. We note that the 3.75-h spin period gives wider failed regions than the 3.5-hr spin period because a transition between compression and tension happens at a spin period of $\sim3.75$ h, causing structurally failed regions to be larger \citep{Hirabayashi2015Sphere}.

\begin{figure}[ht]
  \centering
  \includegraphics{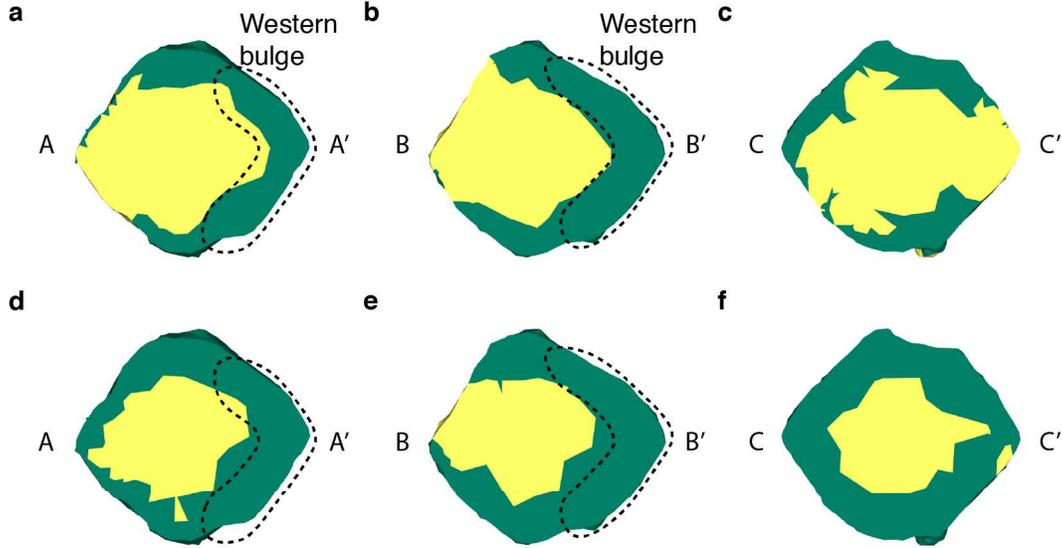}
  \caption{Distributions of the failed regions at spin periods of 3.75 h (Panels a through c) and 3.5 h (Panels d through f). Panels a and d show the failure mode on $AA^\prime$, which is defined in Figure \ref{Fig:top_bottom}. Similarly, Panels b and e indicate the failure mode on $BB^\prime$, and Panels c and f describe that on $CC^\prime$. The dotted lines show the structurally intact regions that appear asymmetrically in the subsurface of the western bulge. Note that Panels c and f are given in \cite{Watanabe2019}.}
  \label{Fig:vertical}  
\end{figure}

We also show the failed regions on the vertical cross sections (Figure \ref{Fig:vertical}). Along $AA^\prime$ and $BB^\prime$, structurally intact regions spread in the subsurface of the western bulge. On the other hand, the distribution of the failed region along $CC^\prime$ is almost symmetric along the spin axis direction. The FEM-derived minimum cohesive strength at spin periods of 3.75 h and 3.5 h are 1.75 Pa and 4 Pa, respectively \citep{Watanabe2019}. Our analysis indicates that the existence of the intact region beneath the western bulge cannot be explained by symmetric deformation. We propose that this results from a deformation process that made the western bulge. 

\section{How fast did Ryugu spin?} 
In the previous section, we argued that the western bulge might result from a deformation process. In this section, we analyze the rotational state that triggered this deformation process. During the deformation occurs at fast spin, fluidized materials with weak mechanical strength \citep{Melosh2011} may reach the equatorial region. If the centrifugal force is strong, such materials should be lofted from there. Because the centrifugal force is proportional to the radial distance from an asteroid's center of mass, this material shedding process can create a circular ridge \citep{Scheeres2006KW4}, which supports the observed circularity of Ryugu \citep{Watanabe2019}. 

To determine this spin period condition, we compute the locations of the dynamical equilibrium points at which the total force is balanced in the rotating frame. If these points touch the surface, fluidized materials should be lofted from there. The force balance condition is described as:
\begin{eqnarray}
{\bfm f}_{g}({\bfm x}) + {\bfm f}_{c}({\bfm x}) = \bfm 0,
\end{eqnarray}
where ${\bfm f}_{g}$ is the gravity force, and ${\bfm f}_{c}$ is the centrifugal force. We compute the gravity force vector by using the algorithm developed by \cite{Werner1997} and a refined shape model that consists of 3,072 facets and 1,538 vertices. These force terms are functions of the dynamical equilibrium points, ${\bfm x}$. We numerically compute the dynamical equilibrium points by using Newton's implicit equation solver \citep[e.g.][]{Press2007}. 

We also compute the zero-velocity curves, which show the pseudo-energy levels that correlate with the locations of the dynamical equilibrium points \citep{Murray2008}. Energetically, these points become either saddle points or local extrema. Seen from Ryugu's center of mass, the region outside the dynamical equilibrium points may experience strong centrifugal forces, while the gravity is dominant in the area within these points. For spinning  top-shaped asteroids, there are usually more than four dynamical equilibrium points because of their axisymmetric shapes \citep{Scheeres2006KW4}. 

\begin{figure}[ht]
  \centering
  \includegraphics{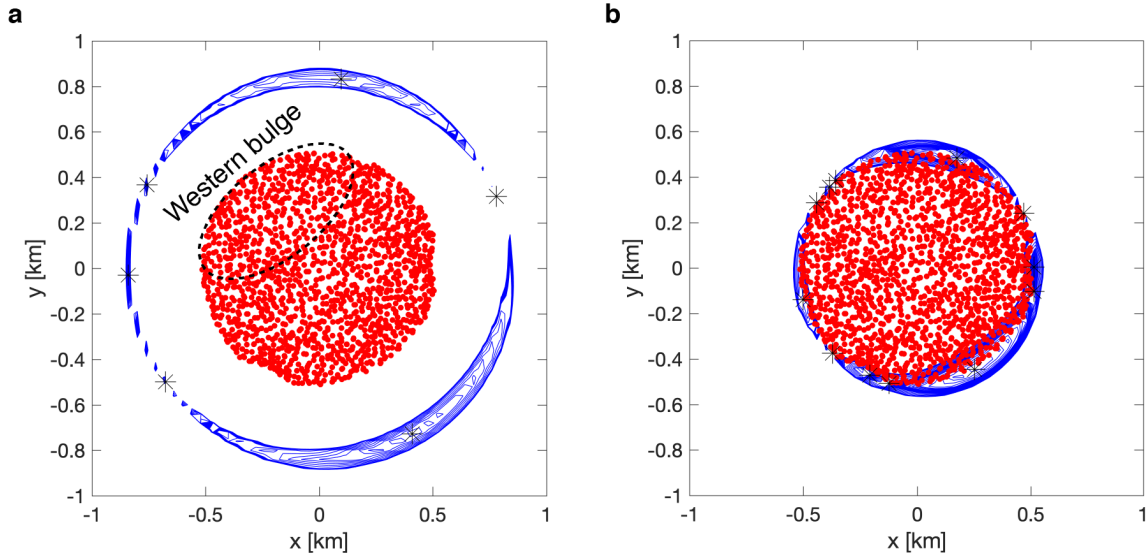}
  \caption{Locations of the dynamical equilibrium points. a shows the case at a spin period of 7.6 h. b describes the case of a spin period of 3.5 h. The red points indicate the shape of Ryugu. The star makers are the locations of the dynamical equilibrium points. The contour outside the shape describes the zero-velocity curves. The red dots describe the shape of Ryugu viewed from the spin axis. The dashed line indicates the location of the western bulge.}
  \label{Fig:EP}  
\end{figure}

Similar to the FEM analysis, we fix the bulk density at 1.2 g cm$^{-3}$. Figure \ref{Fig:EP} shows the locations of the dynamical equilibrium points and the zero velocity curves at spin periods of 7.6 h and 3.5 h. The zero-velocity curves are drawn at pseudo-energy levels near the dynamical equilibrium points to enhance the energy conditions at these points. We find that at a spin period of $3.5$ h, 12 dynamical equilibrium points appear and are about to touch the surface. At this spin condition, the zero velocity curves are circular, the size of which is consistent with Ryujin Dorsum.  

We also give constraints on the cohesive strength of Ryugu. Structural failure should occur in Ryugu at a spin period at 3.5 h if the cohesive strength is $\sim 4$ Pa (see Figure \ref{Fig:failedHour}). This spin period condition may be an upper bound. When Ryugu deforms, its moment of inertia changes due to deformation. If this asteroid rotates at a given spin period before deformation, it should change its spin period after deformation in a manner that the angular momentum is conserved. Because of the rotationally induced deformation mode of a spinning top-shaped asteroid \citep[e.g.][]{Hirabayashi2014DA}, the moment of inertia increases as deformation continues. Thus, before Ryugu's precursor deforms to become the current shape at the 3.5-h spin period, it may have a shorter spin period than that period.

To consider a lower bound of the spin period of deformation, we conduct the following thought experiment. Once Ryugu's precursor structurally failed at a spin period shorter than 3.5 hr, the shape eventually settled into the current configuration at the 3.5-hr spin period. Assuming that it was spinning along the maximum moment of inertia axis, we conservatively choose a sphere that has the smallest moment of inertia to be the shape of the precursor body. Here, the spin period of this object before deformation is considered to be the lower bound. We assume that the total mass and volume are constant, ignoring the amount of shedding materials. Because the mean radius of Ryugu is 448.4 m \citep{Watanabe2019}, we obtain this spin period as 3.0 h and the minimum cohesive strength as $\sim$10 Pa (Figure \ref{Fig:failedHour}). We conclude that the cohesive strength of Ryugu ranges between $\sim 4$ Pa and $\sim 10$ Pa. 

\section{Discussion}
We showed that the subsurface structure of the western bulge is not sensitive to structural failure at present, implying that this region is likely to be relaxed structurally. We proposed that Ryugu experienced a deformation process that generated the western bulge. This process might occur only on the side of the bulge. Such an asymmetric deformation process was numerically predicted for a spheroidal object \citep{Sanchez2016}, who pointed out that even small structural heterogeneity can cause an asymmetric deformation process, although theoretical studies missed this point \citep[e.g.,][]{Holsapple2010, Hirabayashi2015Sphere}. Our hypothesis explains the east-west geomorphological dichotomy. The smooth, less-cratered terrain on the western bulge can be generated due to mass movement. Also, the deformation process can produce void space in the interior, followed by a material infilling process into it, which explains the formation of Tokoyo and Horai Fossae. 

While our discussions were focused on rotationally induced failure, we note alternative possibilities. First, impact craters may contribute to deformation processes in Ryugu. However, we rule out this possibility as a direct contributor because this does not produce the circularity of Ryujin Dorsum. Second, a head-on collision of two similar-sized objects may produce the east-west dichotomy \citep{Leleu2018}. However, although the contact point should be highly compressed and have a ridge-like feature \citep{Leleu2018}, Ryugu does not have such a feature. Instead, we observe depression features, Horai and Tokoyo Fossae. 

We note that our work does not determine whether internal deformation \citep[e.g.][]{Hirabayashi2014DA, Scheeres2016Bennu, Zhang2017, Hirabayashi2018} or surface mass movement without internal deformation \cite[e.g.][]{Walsh2008, Walsh2012, Zhang2017} is the primary cause of the considered asymmetric deformation process. If the internal structure is strong, Ryugu may have surface mass movement. On the other hand, if the interior is uniform, this asteroid may have internal deformation. We, however, rule out large density inhomogeneity between the western bulge and other regions because the center of mass of Ryugu derived by Hayabusa2 corresponds to numerical prediction based on the mission derived shape model with constant density within error \citep{Watanabe2019}. It is also possible that Ryugu may have complex deformation processes if its rubble pile structure possesses internal discontinuities that formed during the reaccumulation process after catastrophic disruption of its parent body. 

\section*{Acknowledgments}
M.H. thanks support from Auburn University and acknowledges ANSYS Mechanical APDL (18.1) under the license owned by Auburn's Samuel Ginn College of Engineering. P.M. acknowledges support from the French space agency CNES as well as from Academies of Excellence: Complex systems and Space, environment, risk and resilience, part of the IDEX JEDI of the Universit{\'e} C{\^o}te d'Azur in connexion with its Center For Planetary Origins. The ONC has been developed under the leadership of JAXA with collaboration with Univ. of Tokyo, Kochi Univ., Rikkyo Univ. Nagoya Univ., Chiba Inst. of Technology, Meiji Univ., Univ. of Aizu, and AIST and contracted contribution of NEC corporation. This study is supported by the JSPS core-to-core program, ``International Network of Planetary Sciences." The FEM results for the 3.5 hr and 3.75 hr periods are from \cite{Watanabe2019} and available at http://hdl.handle.net/11200/49368.

\newpage

\bibliographystyle{aasjournal}


\end{document}